\documentclass[12pt]{article}
\usepackage{amssymb}
\usepackage{color}
\usepackage{graphicx}

\newcommand{\A}{{\mathfrak A}}

\newcommand{\R}{{\cal R}}
\newcommand{\Sc}{{\cal S}}

\newcommand{\mc}{\mathcal}

\newcommand{\be}{\begin{equation}}
\newcommand{\en}{\end{equation}}
\newcommand{\bea}{\begin{eqnarray}}
\newcommand{\ena}{\end{eqnarray}}
\newcommand{\beano}{\begin{eqnarray*}}
\newcommand{\enano}{\end{eqnarray*}}

\newcommand{\1}{1 \!\! 1}
\newcommand{\ST}{\mc S}

\newcommand{\Hil}{\mc H}

\catcode `\@=11 \@addtoreset{equation}{section}
\def\theequation{\arabic{section}.\arabic{equation}}
\catcode `\@=12

\begin{document}

\begin{center}
{\Large \bf Damping in quantum love affairs}   \vspace{2cm}\\

{\large F. Bagarello}
\vspace{3mm}\\
  Dipartimento di Metodi e Modelli Matematici,
Facolt\`a di Ingegneria,\\ Universit\`a di Palermo, I - 90128  Palermo, Italy\\
E-mail: bagarell@unipa.it\\home page:
www.unipa.it$\backslash$\~\,bagarell
\vspace{2mm}\\
\end{center}

\vspace*{2cm}

\begin{abstract}
\noindent In a series of recent papers we have used an operatorial technique to describe stock markets and, in a different context, {\em love affairs} and their time evolutions. The strategy proposed so far does not allow any dumping effect. In this short note we show how, within the same framework, a strictly non periodic or quasi-periodic effect can be introduced in the model by describing in some details a linear Alice-Bob love relation with damping.

\end{abstract}

\vfill

\newpage

\section{Introduction and motivations}

In a series of recent papers, \cite{bag1,bag2,bag3,bag4}, we have introduced and used a {\em number operator} strategy in the description of some simplified models of stock markets. The original reason for adopting this strategy was that all the relevant {\em observables} of the markets (i.e. the quantity which  are used to describe the market) assume  discrete values. The same framework was recently used  for a completely different problem, i.e. for the description of  love affairs, and in particular for a two-actors (Alice and Bob) relation, and a three actors love affair, in which Carla contributes with Alice and Bob to create a love triangle, \cite{fsquare1,fsquare2}. In all these applications the dynamics is generated by a single operator, the {\em hamiltonian} of the system, which is constructed in a very natural way (see below. This, in our opinion, is the most appealing aspect of the procedure) to take into account all the interactions between the different agents of the model. However, the systems considered in \cite{fsquare1,fsquare2} are closed: the actors have no connection with the environment! In particular, this means that Alice can only interact with Bob (and with Carla), but she cannot interact with anyone else. Of course, to be more realistic, this is a point to be reconsidered: Alice's LoA ({\em level of attraction}, see below) for Bob does not only depend on Bob himself, but also on all the other possible interactions that Alice may experience with different people, as well as on different states of mind or other psychological effects. The same holds true for Bob, obviously. In this short note we show how these extra effects can be taken into account within the same general framework we have used so far, and how the presence of the environment introduce the possibility of getting solutions which go down or up in time  to some value fixed by the environment itself. In order to make the details of our approach completely explicit we will discuss here a linear model for which the solution can be deduced analytically, while we postpone the analysis of a non linear model to a paper which is now in preparation, where we will use numerical computations, \cite{fsquare3}.

Of course one may wonder why we should use quantum mechanics in modeling classical systems. The answer, at least as far as our approach is concerned, is that it is quite easy to write down a single operator, the {\em hamiltonian} of the system, which fixes the dynamics of the system: it is enough to follow few simple rules, \cite{bag1}-\cite{fsquare2}. We will come back to this aspect later on. We also want to stress that in recent years quantum mechanical tools have been used more and more in connection with many different classical (complex) systems. We cite here just two recent monographs, \cite{qal,khre}, where other references can be found. More references can be found in \cite{fsquare1}.

\vspace{2mm}

The paper is organized as follows:

In  Section II, after a short review of some results of \cite{fsquare1}, useful to fix the notation, we introduce a new model in which the two actors, Alice and Bob, interact among themselves and with their own environments. We deduce some of the features of this model, in particular the existence of a constant of motion, and we obtain the related equations of motion, working in the Heisenberg picture. We also find the explicit solution for these equations.

In Section III we comment  these results. Then we give our
conclusions and we discuss our plans for the future.

To keep the paper self-contained, we give in the Appendix some basic facts on quantum mechanics and we discuss how an exponential law can be derived by rather general assumptions. Of course, this Appendix is devoted only to those readers which are not familiar with these arguments.

\section{Old and new models}

In \cite{fsquare1} we have introduced a simple model  of two lovers, Bob and Alice, who \emph{interact} exhibiting a certain \emph{interest} for each other. Of course, there are several degrees of possible interest, and to a given Bob's interest  for Alice (LoA, \emph{level of attraction}), there corresponds a related reaction (\emph{i.e.}, a second LoA) of Alice for Bob.

Using a slightly different, and more convenient, notation with respect to that adopted in \cite{fsquare1}, we now introduce $a$ and $b$,  two independent bosonic operators. This means that they obey the commutation rules
\be
[a,a^\dagger]=[b,b^\dagger]=\1,
\label{21}
\en
where $\1$ is the identity operator, while all the other commutators are trivial. Recall that, for two operators $x$ and $y$, $[x,y]=xy-yx$. Here $a$ {\em stands for} (the annihilation operator associated to) Alice and $b$  for (the annihilation operator associated to)  Bob. As in \cite{bag1}-\cite{fsquare2}, bosonic operators are useful since, see Appendix, they {\em create} or {\em annihilate} excitations or, in our language, modify the LoA's of both Alice and Bob. Further, let $\varphi_0$ be the \emph{vacuum} of $a$ and $b$, $a\,\varphi_0=b\varphi_0=0$. If the system is described by $\varphi_0$, then  Bob doesn't experience any attraction for Alice and viceversa, see below. 
Using $\varphi_0$ and the creation operators $a^\dagger$ and $b^\dagger$ we may construct the following vectors:
\be \varphi_{n_a,n_b}:=\frac{1}{\sqrt{n_a!n_b!}}\,(a^\dagger)^{n_a}(b^\dagger)^{n_b}\,\varphi_0,
\label{22}
\en
where $n_j=0,1,2,\ldots$, and $j=a,b$. Let us also define $N_a=a^\dagger\,a$, $N_b=b^\dagger\,b$, and $N=N_a+N_b$.
Hence (see the Appendix), for $j=a, b$,
$N_j\varphi_{n_a,n_b}=n_j\varphi_{n_a,n_b}$, $N_j\,a_j\,\varphi_{n_a,n_b}=(n_j-1)\,a_j\varphi_{n_a,n_b}$, and
$N_j\,a_j^\dagger\,\varphi_{n_a,n_b}=(n_j+1)\,a_j^\dagger\varphi_{n_a,n_b}$.
We also have $N\varphi_{n_a,n_b}=(n_a+n_b)\varphi_{n_a,n_b}$. As usual, the Hilbert space $\Hil$ on which the operators act is obtained by taking the closure of the linear
span of all these vectors, for $n_j\geq0$, $j=a,b$.
A state over the system  is  a normalized linear functional $\omega_{n_a,n_b}$ labeled by two \emph{quantum numbers} $n_a$ and $n_b$ such that $\omega_{n_a,n_b}(x)=\left<\varphi_{n_a,n_b},x\,\varphi_{n_a,n_b}\right>$, where $\left<.,.\right>$ is the scalar product in $\Hil$ and $x$ is an arbitrary operator on $\Hil$. Of course, for generic $x$, $\omega_{n_a,n_b}(x)$ is a complex number; if $x=x^\dagger$, which is what happens in all our computations, $\omega_{n_a,n_b}(x)$ is a real quantity.

In this paper we associate the (integer) eigenvalue  $n_a$ of $N_a$ to the LoA that Alice experiences for Bob: the higher the value of $n_a$ the more Alice desires Bob. For instance, if $n_a=0$, Alice just does not care about Bob. On the other hand, we use $n_b$, the eigenvalue of $N_b$, to label the attraction of Bob for Alice. In \cite{fsquare1} we have assumed that the dynamics of the love affair can be deduced by the following hamiltonian
\be
H=\lambda\left(a^{M}{b^\dagger}+\hbox{h.c.}\right),
\label{23}\en
where $\lambda$ is the interaction parameter, which also play the role of a time scaling, \cite{fsquare1}. Notice that no free hamiltonian is considered here: $H$ consists only of the interaction contribution. The physical meaning of $H$ can be deduced considering the action of, say, $a^{M}\,{b^\dagger}$ on the  vector describing the system at time $t=0$, $\varphi_{n_a,n_b}$. This means that, at $t=0$, Bob is in the state $n_b$, i.e. $n_b$ is Bob's LoA, while Alice is in the state $n_a$.
However, because of the definition of $\varphi_{n_a,n_b}$, $a^{M}\,{b^\dagger}\,\varphi_{n_a,n_b}$, which is different from zero only if $M<n_a$,
 is proportional to $\varphi_{n_a-M,n_b+1}$. Hence, Alice's interest for Bob decreases of $M$  units while Bob's interest for Alice increases of 1 unit. Of course,  because of the presence of $b\,{a^\dagger}^M$ in $H$, if $n_b\geq1$ we see that
$b\,{a^\dagger}^M\varphi_{n_a,n_b}$ is proportional to $\varphi_{n_a+M,n_b-1}$: hence, Alice's interest for Bob is increasing (of $M$ units) while Bob looses interest in Alice. Hence $H$ describes quite well what in \cite{fsquare1} has been assumed to be a natural law in love affairs: the more Bob is attracted by Alice, the less Alice cares about Bob, and viceversa. $I(t):=N_a(t)+M\, N_b(t)$ is a constant of motion: $I(t)=I(0)=N_a(0)+M\, N_b(0)$, for all $t\geq0$: $[H,I]=0$. Therefore, during the time evolution, a certain {\em global attraction} is preserved and can only be exchanged between Alice and Bob. Notice that the existence of such a constant of motion is in agreement with our naive vision of a love affair sketched above.

The hamiltonian (\ref{23}) produces  the following differential system for the
annihilation operators $a(t)$ and $b(t)$:
\be
\begin{array}{ll} \dot a(t)=-i\,\lambda M\,b(t)(a(t)^\dagger)^{M-1}\\
\dot b(t)=-i\,\lambda\,(a(t))^M.\end{array}
\label{25}
\en
Then, we may use the solutions of these equations to construct $N_a(t):=a^\dagger(t)\,a(t)$ and $N_b(t):=b^\dagger(t)\,b(t)$. The operators $a(t)$ and $b(t)$ can be found analytically if $M=1$. In this case, which corresponds to the assumption that Alice and Bob react in the same identical way (but for a sign), the solution of system (\ref{25}) is
\be
a(t)=a\cos(\lambda t)-ib\sin(\lambda t), \qquad
b(t)=b\cos(\lambda t)-i a \sin(\lambda t).
\en
Now, if we assume that at $t=0$ Bob and Alice are respectively in the  $n_a$'th and $n_b$'th LoA's, the state of the system at $t=0$ is $\omega_{n_a,n_b}$.  Therefore, calling $n_j(t):=\omega_{n_a,n_b}(N_j(t))$, $j=a,b$, we find that
\be\label{n1lin2}
n_a(t)=n_a\cos^2(\lambda t)+n_b\sin^2(\lambda t),\qquad n_b(t)=n_b\cos^2(\lambda t)+n_a\sin^2(\lambda t).
\en
 Hence $\omega_{n_a,n_b}(I(t))=n_a+n_b$, as expected. The conclusion is quite simple and close to our view of how the law of the attraction should work: the infatuations of Alice and Bob oscillate in such a way that when Bob's LoA increases, that of Alice decreases and viceversa, with a period which is directly related to the value of the interaction parameter $\lambda$, which therefore can be seen as a time scaling. In particular, as it is natural, if $\lambda=0$  equation (\ref{n1lin2}) shows that both Alice and Bob stay in their initial LoA's.

Much harder is the situation when $M>1$, \cite{fsquare1}, for which a numerical approach seems more appropriate. Here we are interested in introducing within our scheme the role of the environment of both Alice and Bob, in a such a way that the two main actors of our game can get in touch with the external world. We believe that, doing so, the model approaches more and more a real system. However, since here we are  interested in describing the general ideas of our method, we will only consider  linear interactions, i.e. we fix $M=1$. In this way an analytical solution will be found and the general strategy will appear clearly with no extra difficulties arising from non linearity. In a paper in preparation, \cite{fsquare3}, non linear effect will also be considered and a numerical approach will be proposed.

\subsection{Enriching the model}

Suppose now that Alice and Bob interact with their own reservoirs. We call $\Sc=\Sc_a\cup\Sc_b\cup\R_A\cup\R_B$ the full system, made of Alice ($\Sc_a$), Bob ($\Sc_b$), Alice's reservoir ($\R_A$), and Bob's reservoir ($\R_B$). The hamiltonian for $\Sc$ looks now like
\be\left\{
\begin{array}{ll}
H=H_A+H_B+\lambda H_I,\\
H_A=\omega_aa^\dagger a+\int_{\Bbb{R}}\Omega_A(k)A^\dagger(k)A(k)\,dk+\gamma_A\int_{\Bbb{R}}\left(a^\dagger A(k)+a A^\dagger(k)\right)\,dk,\\
H_B=\omega_bb^\dagger b+\int_{\Bbb{R}}\Omega_B(k)B^\dagger(k)B(k)\,dk+\gamma_B\int_{\Bbb{R}}\left(b^\dagger B(k)+b B^\dagger(k)\right)\,dk,\\
H_I=a^\dagger b+a b^\dagger.
\label{26}\end{array}
\right.
\en
All the constant in (\ref{26}) are real quantities. It is worth mentioning that a regularization could be also considered in the definition above to make the hamiltonian rigorously defined. We will skip this mathematical  details here, since we are more interested in the physical meaning of $H$ and since, using for instance the {\em stochastic limit} approach, \cite{accbook}, we can make our treatment rigorous. The following bosonic commutation rules are assumed:
\be
[a,a^\dagger]=[b,b^\dagger]=\1,\quad [A(k),A^\dagger(q)]=[B(k),B^\dagger(q)]=\1\delta(k-q),
\label{27}\en
while all the other commutators are zero. $H_A$ and $H_B$ respectively describe the interaction of Alice and Bob with their own reservoirs, which consist of several (infinite) ingredients. Their forms are, in a sense, standard for systems interacting with a reservoir, see \cite{barrad} for instance. These can be interpreted as other men, other women, or as the effect of the tiredness or tediousness in the sentimental relation, as well as some other (psychological) effects. $H_I$ contains our (linear) interaction between Alice and Bob, which follows the same rules of our previous model: for instance, $a^\dagger\,b$ describes an increasing Bob's LoA  and a simultaneously decreasing Alice's LoA.

The Heisenberg equations of motion for the annihilation operators are the following
\be\left\{
\begin{array}{ll}
\dot a(t)=i[H,a(t)]=-i\omega_a a(t)-i\gamma_A\int_{\Bbb{R}}A(k,t)\,dk-i\lambda b(t),\\
\dot b(t)=i[H,b(t)]=-i\omega_b b(t)-i\gamma_B\int_{\Bbb{R}}B(k,t)\,dk-i\lambda a(t),\\
\dot A(k,t)=i[H,A(k,t)]=-i\Omega_A(k) A(k,t)-i\gamma_A a(t),\\
\dot B(k,t)=i[H,B(k,t)]=-i\Omega_B(k) B(k,t)-i\gamma_B b(t).
\label{28}\end{array}
\right.
\en
Using the same strategy discussed in Appendix, i.e. choosing $\Omega_A(k)=\Omega_A k$ and $\Omega_B(k)=\Omega_B k$, $\Omega_A,\Omega_B>0$, we further get
\be\left\{
\begin{array}{ll}
\dot a(t)=-\nu_A a(t)-i\lambda b(t)-i\gamma_A f_A(t),\\
\dot b(t)=-\nu_B b(t)-i\lambda a(t)-i\gamma_B f_B(t),
\label{29}\end{array}
\right.
\en
where $\nu_A=\frac{\pi\gamma_A^2}{\Omega_A}$, $\nu_B=\frac{\pi\gamma_B^2}{\Omega_B}$, $f_A(t)=\int_{\Bbb{R}}e^{-i\Omega_A k t}A(k)\,dk$ and $f_B(t)=\int_{\Bbb{R}}e^{-i\Omega_B k t}B(k)\,dk$.  From (\ref{29}) we deduce the following second order differential equation for $a(t)$:
\be
\ddot a(t)+\dot a(t)(\nu_A+\nu_B)+a(t)(\nu_A\nu_A+\lambda^2)=\Phi(t),
\label{210}\en
where $\Phi(t):=-i\gamma_A \dot f_A(t)-i\nu_B\gamma_A f_A(t)-\lambda \gamma_B f_B(t)$. To find the solution of this equation, satisfying $a(0)=a$ and $b(0)=b$, we need to compute the Green function for the differential operators $L:=\frac{d^2}{dt^2}+(\nu_A+\nu_B)\,\frac{d}{dt} +(\nu_A\nu_A+\lambda^2)$, i.e. the function $G(t)$ satisfying $L[G](t)=\delta(t)$. To keep the computations reasonably simple from now on we fix $\nu_A=\nu_B=:\nu$, which implies that $\frac{\gamma_A^2}{\Omega_A}=\frac{\gamma_B^2}{\Omega_B}$, and $\omega_A=\omega_B=:\omega$. It is a standard exercise in Fourier transform to deduce that
$$
G(t)=\left\{
\begin{array}{ll}
\frac{1}{\lambda}\,\sin(\lambda t)\,e^{-\nu t},\qquad t>0\\
0, \hspace{3cm}\mbox{otherwise}.
\label{211}\end{array}
\right.
$$
The general solution of the equation $L[a_0(t)]=0$ is $a_0(t)=x_+e^{\epsilon_+t}+x_-e^{\epsilon_-t}$, with $\epsilon_\pm=\left(-\nu\pm i\lambda\right)$. After few computations and recalling that $b(t)$ can be deduced from $a(t)$ by rewriting (\ref{29}) as $b(t)=\frac{1}{\lambda}\left(i\dot a(t)+i\nu_A a(t)-\gamma_A f_A(t)\right)$, we get
\be
\left\{
\begin{array}{ll}
 a(t)=a e^{-\nu t}\cos(\lambda t)-ibe^{-\nu t}\sin(\lambda t)+R_a(t),\\
 b(t)=b e^{-\nu t}\cos(\lambda t)-iae^{-\nu t}\sin(\lambda t)+R_b(t),
\label{212}\end{array}
\right.
\en
where we have defined the following functions:
\be
\left\{
\begin{array}{ll}
R_a(t)=\rho(t)+\frac{e^{-\nu t}}{\lambda}\left\{i\Gamma(0)\sin(\lambda t)-\lambda \rho(0)\cos(\lambda t)\right\},\\
R_b(t)=\frac{1}{\lambda}\left\{\Gamma(t)-\Gamma(0)e^{-\nu t}\cos(\lambda t)\right\}+i \rho(0)e^{-\nu t}\sin(\lambda t),\\
\rho(t)=\int_{\Bbb{R}}\left(\rho_A(k,t)A(k)+\rho_B(k,t)B(k)\right)\,dk,\\
\Gamma(t)=\int_{\Bbb{R}}\left(\Gamma_A(k,t)A(k)+\Gamma_B(k,t)B(k)\right)\,dk,
\label{213}\end{array}
\right.
\en
as well as $$\left\{
\begin{array}{ll}\rho_A(k,t)=-\gamma_A\,\frac{\Omega_A k-i\nu }{\lambda^2+(i\Omega_A k-\nu)^2}\,e^{-i\Omega_Akt},\qquad \rho_B(k,t)=\frac{-\lambda\gamma_B }{\lambda^2+(i\Omega_B k-\nu)^2}\,e^{-i\Omega_Bkt},\\ \Gamma_A(k,t)=i\dot\rho_A(k,t)+i\nu\rho_A(k,t)-\gamma_Ae^{-i\Omega_Akt},\qquad
\Gamma_B(k,t)=i\dot\rho_B(k,t)+i\nu\rho_B(k,t).\end{array}
\right.$$
It is now easy to find the mean value of the number operators $N_a(t)=a^\dagger(t)a(t)$ and $N_b(t)=b^\dagger(t)b(t)$ on a state over $\Sc$, which is of the form $\omega_\Sc(.):=\left<\varphi_{n_a,n_b},\,.\,\varphi_{n_a,n_b}\right>\omega_\R(.)$. Here $\omega_\R$ is a suitable state over the reservoir $\R=\R_A\cup\R_B$, while, as usual, $\varphi_{n_a,n_b}$ is the eigenstate of $N_a$ and $N_b$ with eigenvalues $n_a$ and $n_b$. Then, calling $n_a(t)=\omega_\Sc(N_a(t))$ and $n_b(t)=\omega_\Sc(N_b(t))$, and assuming that, see Appendix,
\be
\omega_\R(A^\dagger(k)A(q))=N_A(k)\delta(k-q), \quad \omega_\R(B^\dagger(k)B(q))=N_B(k)\delta(k-q),
\label{214}\en
$N_A(k)$ and $N_B(k)$ to be fixed, we conclude that
$$
n_a(t)=e^{-2\pi\gamma_A^2 t/\Omega_A}\left(n_a\cos^2(\lambda t)+n_b\,\sin^2(\lambda t)\right)+$$ \be+
\int_{\Bbb{R}}\left(N_A(k)|\mu_{a,A}(k,t)|^2+N_B(k)|\mu_{a,B}(k,t)|^2\right)\,dk
\label{215}\en
and
$$
n_b(t)=n_b\,e^{-2\pi\gamma_A^2 t/\Omega_A}\,\cos^2(\lambda t)+n_a\,e^{-2\pi\gamma_A^2 t/\Omega_A}\,\sin^2(\lambda t)+$$ \be+
\int_{\Bbb{R}}\left(N_A(k)|\mu_{b,A}(k,t)|^2+N_B(k)|\mu_{b,B}(k,t)|^2\right)\,dk,
\label{216}\en
where we have introduced the following functions:
\beano
\mu_{a,A}(k,t)=\rho_A(k,t)+\frac{i}{\lambda}e^{-\nu t}\sin(\lambda t)\Gamma_A(k,0)-e^{-\nu t}\cos(\lambda t)\rho_A(k,0),\\
\mu_{a,B}(k,t)=\rho_B(k,t)+\frac{i}{\lambda}e^{-\nu t}\sin(\lambda t)\Gamma_B(k,0)-e^{-\nu t}\cos(\lambda t)\rho_B(k,0),\\
\mu_{b,A}(k,t)=\frac{1}{\lambda}\,\Gamma_A(k,t)-\frac{1}{\lambda}e^{-\nu t}\cos(\lambda t)\Gamma_A(k,0)-e^{-\nu t}\sin(\lambda t)\rho_A(k,0),\\
\mu_{b,B}(k,t)=\frac{1}{\lambda}\,\Gamma_B(k,t)-\frac{1}{\lambda}e^{-\nu t}\cos(\lambda t)\Gamma_B(k,0)-e^{-\nu t}\sin(\lambda t)\rho_B(k,0).
\enano
Formulas (\ref{215}) and (\ref{216}) look interesting since they clearly display the different contributions arising from the system $\Sc_a\cup\Sc_b$ and from the reservoir $\R$. Of course, different choices of the functions of the reservoir $N_A(k)$ and $N_B(k)$ clearly produces different expressions for the LoA's of Alice and Bob. In particular, if for instance $N_A(k)=N_B(k)=0$ almost everywhere in $k$, we get
$$
n_a(t)=e^{-2\pi\gamma_A^2 t/\Omega_A}\left(n_a\cos^2(\lambda t)+n_b\,\sin^2(\lambda t)\right),$$
\be
n_b(t)=e^{-2\pi\gamma_A^2 t/\Omega_A}\left(n_b\cos^2(\lambda t)+n_a\,\sin^2(\lambda t)\right),
\label{217}\en
which produce \underline{damped} oscillations for both Alice and Bob: independently of their initial status, the effect of the reservoirs is to switch off the love between Alice and Bob, at least for this trivial choice of $N_A(k)$ and $N_B(k)$. In this case, if for instance,  $n_a=n_b$, i.e. if  Alice and Bob experience the same LoA at $t=0$, then we get $n_a(t)=n_a\,e^{-2\pi\gamma_A^2 t/\Omega_A}$ and $n_b(t)=n_b\,e^{-2\pi\gamma_A^2 t/\Omega_A}$. The speed of decay of their LoA is related to $\gamma_A^2/\Omega_A$ which, we recall, coincides with  $\gamma_B^2/\Omega_B$. In particular, the stronger the interaction between, say, Alice and her reservoir, the faster the decay to zero of her love for Bob. Of course, a different speed (i.e. behavior) for Alice and Bob is expected, in general. This can be recovered assuming that $\nu_A\neq\nu_B$. This situation will be considered in \cite{fsquare3}.



\section{Final comments}

\begin{figure}[h]
\begin{center}
\includegraphics[width=0.47\textwidth]{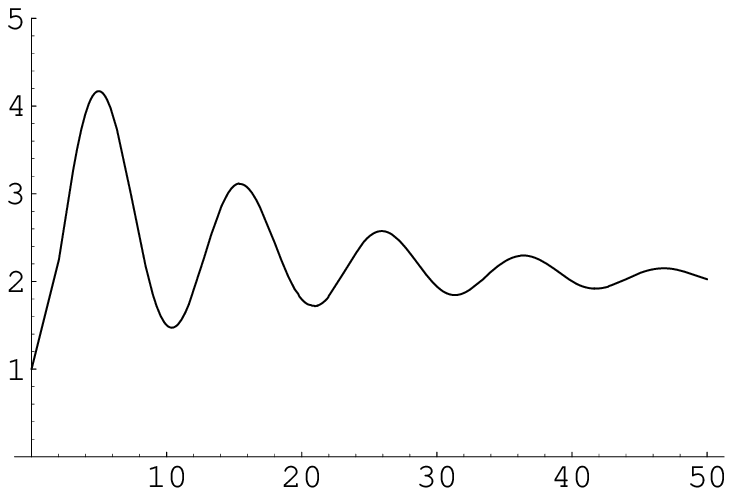}\hspace{8mm}
\includegraphics[width=0.47\textwidth] {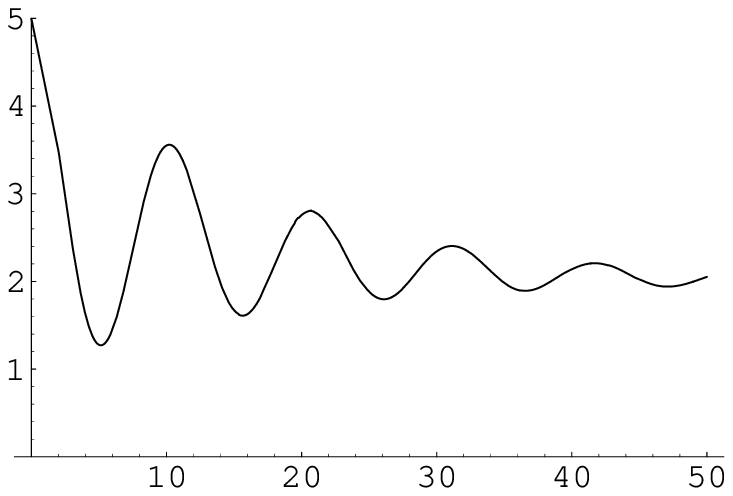}\hfill\\
\caption{\label{fig1}\footnotesize $n_a(t)$ (left) and $n_b(t)$ (right), for $n_a=1$ and $n_b=5$}
\end{center}
\end{figure}

Let us now consider what happens if $N_A(k)$ and $N_B(k)$ are different from zero. First we consider the following case: $N_A(k)=N_B(k)=5\,e^{-k^2}$. Here 5 is chosen because, just to fix the ideas, we will consider in the following that, at $t=0$, the initial conditions on $n_a$ and $n_b$ are simply 1 or 5: $n_a=1$ and $n_b=5$, see Figure \ref{fig1}, corresponds to a low Alice's LoA and an high Bob's LoA. The other parameters are chosen to be the following: $\omega=1$, $\Omega_A=\Omega_B=1$, $\gamma_A=\gamma_B=.1$ and $\lambda=.3$. So the situation looks rather symmetrical and the only asymmetry is given by the initial conditions above.


These figures show a non purely oscillatory behaviors of $n_a(t)$ and $n_b(t)$, which seem to converge to a limiting asymptotic value for large $t$: this appears to be a sort of intermediate state (a sort of attractor) in which both Bob and Alice will continue their relationship with not many changes in their LoA'a. However, different situations may arise. First of all, if we just change the initial conditions requiring that $n_a=n_b=5$, it is easy to see that $n_a(t)=n_b(t)$ and that they both decay monotonically to a value close to 2.5.

\begin{figure}[h]
\begin{center}
\includegraphics[width=0.47\textwidth]{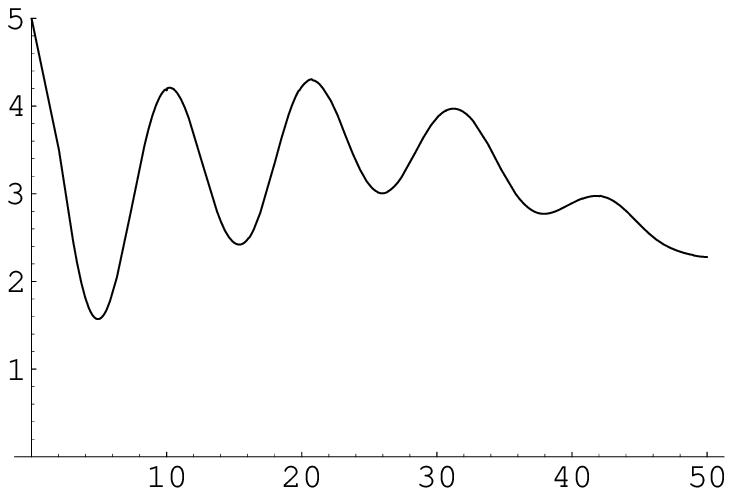}\hspace{8mm}
\includegraphics[width=0.47\textwidth] {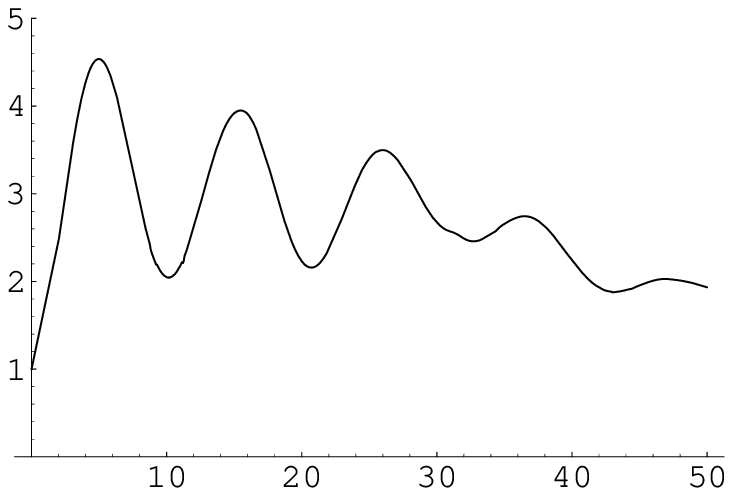}\hfill\\
\caption{\label{fig2}\footnotesize $n_a(t)$ (left) and $n_b(t)$ (right),  for $n_a=5$ and $n_b=1$}
\end{center}
\end{figure}

Moreover, if we now break down the original symmetry between Alice and Bob not only considering different initial conditions but also changing the values of the parameters, then the asymptotic behavior of $n_a(t)$ and $n_b(t)$ is not so clear, see Figure \ref{fig2}. This is obtained taking $\omega$ and $\lambda$ as above, and $\Omega_A=1$, $\Omega_B=4$, $\gamma_A=.1$ and $\gamma_B=.2$. This choice satisfies the requirement $\gamma_A^2/\Omega_A=\gamma_B^2/\Omega_B$.

\begin{figure}
\begin{center}
\includegraphics[width=0.47\textwidth]{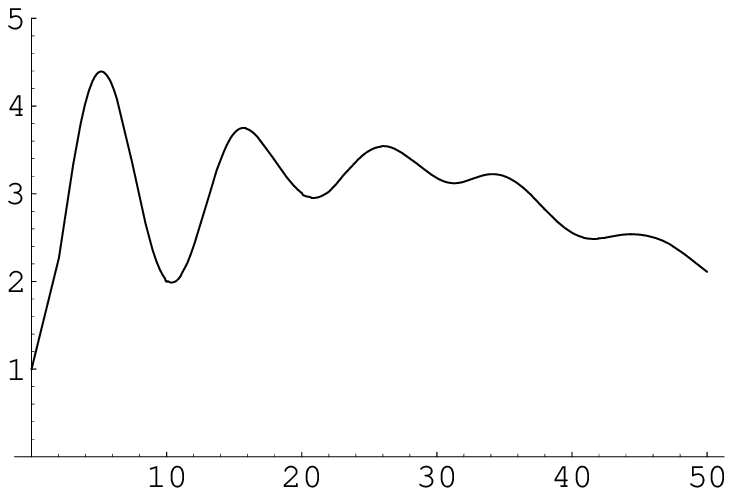}\hspace{8mm}
\includegraphics[width=0.47\textwidth] {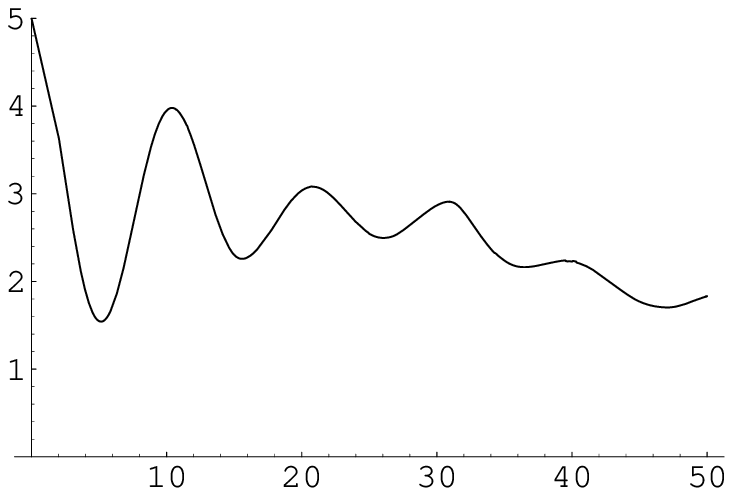}\hfill\\
\caption{\label{fig3}\footnotesize $n_a(t)$ (left) and $n_b(t)$ (right),  for $n_a=1$ and $n_b=5$}
\end{center}
\end{figure}

\begin{figure}
\begin{center}
\includegraphics[width=0.47\textwidth]{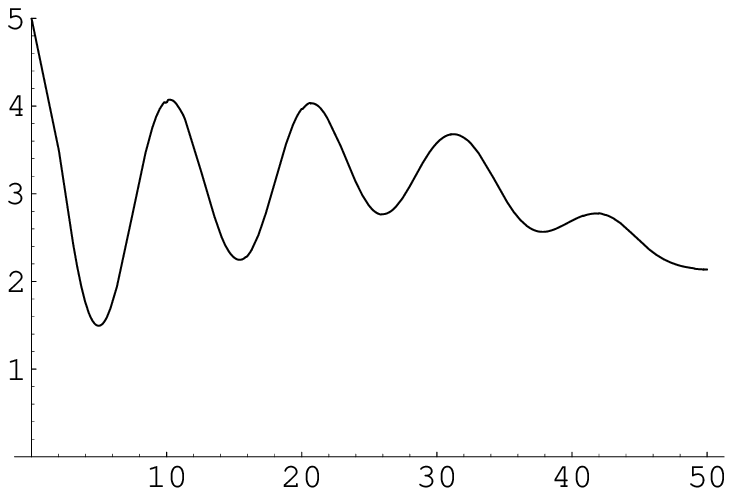}\hspace{8mm}
\includegraphics[width=0.47\textwidth] {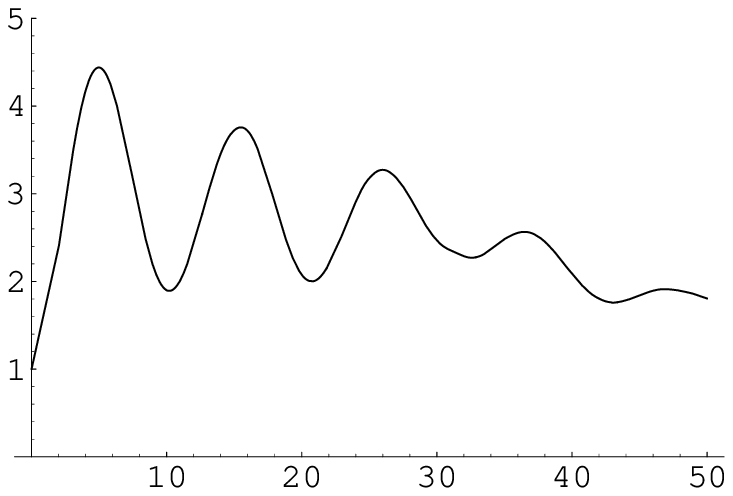}\hfill\\
\caption{\label{fig4}\footnotesize $n_a(t)$ (left) and $n_b(t)$ (right),  for $n_a=5$ and $n_b=1$}
\end{center}
\end{figure}

\begin{figure}
\begin{center}
\includegraphics[width=0.47\textwidth]{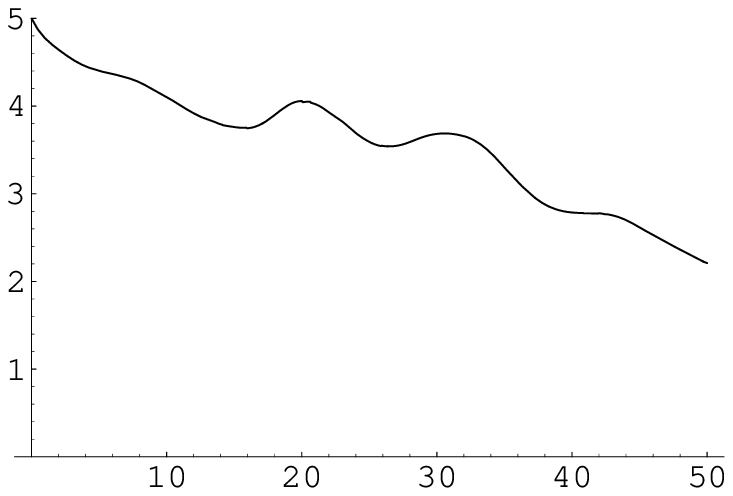}\hspace{8mm}
\includegraphics[width=0.47\textwidth] {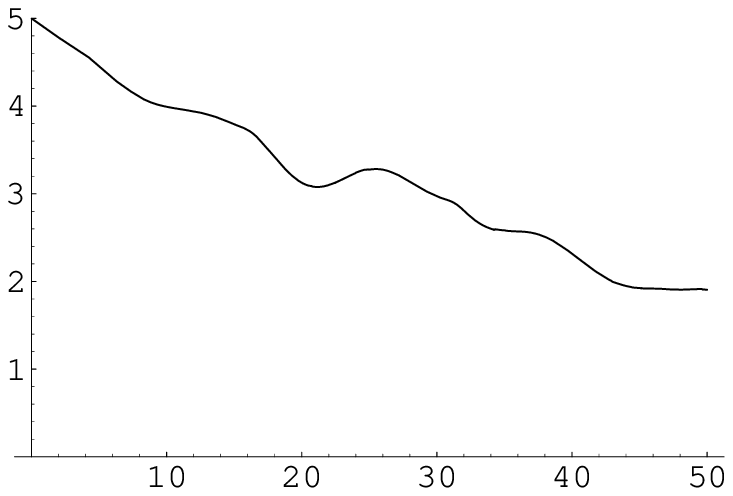}\hfill\\
\caption{\label{fig5}\footnotesize $n_a(t)$ (left) and $n_b(t)$ (right),  for $n_a=5$ and $n_b=5$}
\end{center}
\end{figure}

We see from the figure that $n_a(t)$ and $n_b(t)$ do not apparently tends to a given value for $t$ increasing.  Similar plots are obtained if we increase even more the asymmetry between Alice and Bob by considering their interactions with two different reservoirs. Indeed, if we take now $N_A(k)=5\,e^{-k^2}$ and $N_B(k)=5\,e^{-3k^2}$, leaving all the other constants unchanged, we get the functions plotted in Figure \ref{fig3}, for $n_a=1$ and $n_b=5$, in Figure \ref{fig4}, for $n_a=5$ and $n_b=1$, and in Figure \ref{fig5}, for $n_a=5$ and $n_b=5$.

In particular Figure \ref{fig5} shows that $n_a(t)$ and $n_b(t)$ both decrease, not monotonically, and that some limiting point, if any, is reached very far away (compared with Figure \ref{fig1}) in time. Moreover, this same figure also clearly displays that $n_a(t)+n_b(t)$ is not a constant of motion, as expected since the contribution of the reservoirs cannot be neglected. In fact, we can show that an integral of motion still exists for this open system, and it looks like $J=N_a+N_b+N_A+N_B$, where $N_a=a^\dagger\,a$, $N_b=b^\dagger\,b$, $N_A=\int_{\Bbb{R}}A^\dagger(k)A(k)\,dk$ and $N_B=\int_{\Bbb{R}}B^\dagger(k)B(k)\,dk$: $[J,H]=0$. This implies that the oscillations and the decay of the functions $n_a(t)$ and $n_b(t)$ must be related to some oscillations and to an overall increasing behavior of $N_A(t)$ and $N_B(t)$. Therefore, Alice's LoA decreases because she interacts with other sources. The same happens to Bob. However, if they are lucky enough, their LoA's both oscillate around some strictly positive value, as in Figure \ref{fig1} for instance. So their love relation can survive for a long time, even if with less {\em strength} that at the beginning. This is realistic, indeed. If they are not lucky, see formulas (\ref{217}), their love just vanishes after some oscillations. Our analysis shows that, to be concrete, their {\em luck} really depends on the parameters of their reservoirs.

\vspace{4mm}

In this paper we have continued our analysis of sentimental relationships using operatorial techniques, first discussed in \cite{fsquare1}. In our opinion, our previous models have been made significantly more realistic by adding the possibility of damping effects which were not there in the original scheme. These effects are related, as we have seen, to the possibilities for Alice and Bob to interact with the outer world, i.e. with  friends, other lovers, parents and relatives for instance, which in a realistic love affair play a relevant role in the evolution of the relation.

This improvement will be used in the forthcoming construction of even more realistic (probably non-linear) models and in the analysis of other biological and economical systems, for which, however, producing exactly solvable models is much more difficult.

\section*{Acknowledgements}

This work has been financially supported in part by M.U.R.S.T.,
within the  project {\em Problemi Matematici Non Lineari di
Propagazione e Stabilit\`a nei Modelli del Continuo}, coordinated
by Prof. T. Ruggeri.

\vspace{8mm}

 \appendix

\renewcommand{\theequation}{\Alph{section}.\arabic{equation}}

 \section{\hspace{-.7cm}ppendix: Some general results}

 \subsection{Mathematical background}

Let $\Hil$ be an Hilbert space, $\ST$ our physical system and $\A$ the set of
all the operators on $\Hil$ useful for a complete description of $\ST$
(the {\em observables } of $\ST$). In many physical systems $\A$ consists also of unbounded operators. This creates mathematical difficulties which will not be considered here, since the relevant part of their spectra are bounded. The
description of the time evolution of $\ST$ is related to a
self-adjoint operator $H=H^\dagger$, which will be assumed not to
depend explicitly on time, which is called {\em the hamiltonian}
of $\ST$. In the {\em Heisenberg} representation
the time evolution of an observable $X\in\A$ is given by $
X(t)=e^{iHt}Xe^{-iHt}$ or, equivalently, by the
solution of the differential equation $
\frac{dX(t)}{dt}=ie^{iHt}[H,X]e^{-iHt}=i[H,X(t)],$
where $[A,B]:=AB-BA$ is the {\em commutator } between $A$ and $B$.

In this paper a special role is played by the so called {\em
canonical commutation relations } (CCR): we say that a set of
operators $\{a_l,\,a_l^\dagger, l=1,2,\ldots,L\}$ satisfy the CCR
if the following hold:\be
[a_l,a_n^\dagger]=\delta_{ln}\1,\hspace{8mm}
[a_l,a_n]=[a_l^\dagger,a_n^\dagger]=0, \label{sr1}\en for all
$l,n=1,2,\ldots,L$, see \cite{mer}. The operators $\hat n_l=a_l^\dagger a_l$ and $\hat
N=\sum_{l=1}^L \hat n_l$ are both self-adjoint operators. In
particular $\hat n_l$ is the {\em number operator} for the l-th
mode, while $\hat N$ is the {\em number operator of $\ST$}.

The Hilbert space of our system is constructed as follows: we
introduce the {\em vacuum} of the theory, that is a vector
$\varphi_0$ which is annihiled by all the {\em annihilation}
operators $a_l$: $a_l\varphi_0=0$, for $l=1,2,\ldots,L$. Then
we act on $\varphi_0$ with the {\em creation} operators
$a_l^\dagger$: \be
\varphi_{n_1,n_2,\ldots,n_L}:=\frac{1}{\sqrt{n_1!\,n_2!\ldots
n_L!}}(a_1^\dagger)^{n_1}(a_2^\dagger)^{n_2}\cdots
(a_L^\dagger)^{n_L}\varphi_0 \label{sr2}\en These vectors form an
orthonormal set and are eigenstates of both $\hat n_l$ and $\hat
N$: $\hat
n_l\varphi_{n_1,n_2,\ldots,n_L}=n_l\varphi_{n_1,n_2,\ldots,n_L}$
and $\hat
N\varphi_{n_1,n_2,\ldots,n_L}=N\varphi_{n_1,n_2,\ldots,n_L}$,
where $N=\sum_{l=1}^Ln_l$. Moreover we have $\hat n_l
(a_l\varphi_{n_1,n_2,\ldots,n_L})=(n_l-1)(a_l\varphi_{n_1,n_2,\ldots,n_L})$
and $\hat n_l
(a_l^\dagger\varphi_{n_1,n_2,\ldots,n_L})=(n_l+1)(a_l^\dagger\varphi_{n_1,n_2,\ldots,n_L})$.
The Hilbert space is obtained by taking the closure of the linear
span of all these vectors.

An operator $Z\in\A$ is a {\em constant of motion} if it commutes
with $H$. Indeed in this case the equation of motion for $Z(t)$ reduces to
$\dot Z(t)=0$, so that $Z(t)=Z$ for all $t$.

The vector $\varphi_{n_1,n_2,\ldots,n_L}$ in (\ref{sr2}) defines a
{\em vector (or number) state } over the algebra $\A$  as
\be\omega_{n_1,n_2,\ldots,n_L}(X)=
<\varphi_{n_1,n_2,\ldots,n_L},X\varphi_{n_1,n_2,\ldots,n_L}>,\label{sr3}\en
where $<\,,\,>$ is the scalar product in  $\Hil$. We refer to \cite{bag1} for further details on this subject.

\subsection{An exponential law}

Suppose now that we have a first system, $\ST$, interacting with a second system, $\tilde\ST$, and suppose that both $\ST$ and $\tilde\ST$ are {\em of the same size}: by this we mean that $\ST$ describes a single actor whose related (bosonic) operators are $a$, $a^\dagger$ and $\hat n_a= a^\dagger a$ and, analogously, $\tilde\ST$ describes a second actor whose related (again, bosonic) operators are $b$, $b^\dagger$ and $\hat n_b= b^\dagger b$. These operators obey the following rules: $[a,a^\dagger]=[b,b^\dagger]=\1$, while all the other commutators are zero. A natural choice for the hamiltonian of the two coupled systems is $h=\omega_a\hat n_a+\omega_b\hat n_b+\mu\left(a^\dagger b+b^\dagger a\right)$, where $\omega_a, \omega_b$ and $\mu$ are real quantities, in order to have $h=h^\dagger$. $h$ contains a free part plus an interaction which is such that, if the eigenvalue of $\hat n_a$ increases of one unit during the time evolution, then the eigenvalue of $\hat n_b$ must decreases of one unit, and viceversa. Moreover  $[h,\hat n_a+\hat n_b]=0$, so that $\hat n_a+\hat n_b$ is an integral of motion. The equations of motion for $a(t)$ and $b(t)$ can be deduced as follows
$$
\dot a(t)=i[h,a(t)]=-i\omega_a a(t)-i\mu b(t),\qquad  \dot b(t)=i[h,b(t)]=-i\omega_b b(t)-i\mu a(t),
$$
whose solution can be written as $a(t)=\alpha_a(t)\,a+\alpha_b(t)\,b$ and $b(t)=\beta_a(t)\,a+\beta_b(t)\,b$, where the functions $\alpha_j(t)$ and $\beta_j(t)$, $j=a,b$, are linear combinations of $e^{\lambda_\pm t}$, with $\lambda_\pm=\frac{-i}{2}(\omega_a+\omega_b-\sqrt{(\omega_a-\omega_b)^2+4\mu^2})$. Moreover $\alpha_a(0)=\beta_b(0)=1$ and $\alpha_b(0)=\beta_a(0)=0$, so that $a(0)=a$ and $b(0)=0$. Hence we see that both $a(t)$ and $b(t)$, and $\hat n_a(t)$ and $\hat n_b(t)$ as a consequence, are linear combinations of oscillating functions,: no damping is possible within this simple model.

\vspace{2mm}

Suppose now that the system $\tilde\ST$ is replaced by an infinitely extended reservoir $\R$, whose actors are described by an infinite set of bosonic operators $b(k), b^\dagger(k)$ and $\hat n(k)=b^\dagger(k) b(k)$, $k\in\Bbb{R}$. The hamiltonian of $\ST$ plus $\R$ which extends $h$ above is
\be
H=H_0+\lambda H_I,\qquad H_0=\omega \hat n_a+\int_{\Bbb{R}}\omega(k)\hat n(k)\,dk,\quad H_I=\int_{\Bbb{R}}\left(a b^\dagger(k)+a^\dagger b(k)\right)f(k) dk,
\label{sr4}\en
where $[a,a^\dagger]=\1$, $[b(k),b^\dagger(q)]=\1\delta(k-q)$, while all the other commutators are zero. All the constant appearing in (\ref{sr4}), as well as the regularizing function $f(k)$, are real, so that $H=H^\dagger$. At a certain point in this appendix we will take $f(k)=1$: this will make easier the computation of the solution of the equations of motion, but makes our $H$ a formal object. A more rigorous approach, which we will not consider here, can be settled up using the results in \cite{accbook}. Notice that an integral of motion again exists, ${\mathfrak N}:=\hat n_a+\int_{\Bbb{R}}\hat n(k)\,dk$, which extends the previous one.

With this choice of $H$ the equation of motions are
\be\label{sr5}\left\{
\begin{array}{ll}
\dot a(t)=i[H,a(t)]=-i\omega a(t)-i\lambda \int_{\Bbb{R}} f(k)\,b(k,t)\,dk,\\
\dot b(k,t)=i[H,b(k,t)]=-i\omega(k) b(k,t)-i\lambda f(k)\,a(t),\end{array}
\right.\en
which are supplemented by the initial conditions $a(0)=a$ and $b(k,0)=b(k)$.
The second equation in (\ref{sr5}) can be rewritten as $b(k,t)=b(k)e^{-i\omega(k)t}-i\lambda f(k)\int_0^t a(t_1)e^{-i\omega(k)(t-t_1)}\,dt_1$. Fixing now $f(k)=1$, assuming that $\omega(k)=k$, and replacing $b(k,t)$ in the first equation in (\ref{sr5}), we find that
$$
\dot a(t)=-(i\omega+\pi\lambda^2) a(t)-i\lambda \int_{\Bbb{R}} b(k)\,e^{-ikt}\,dk.
$$
Here we have also changed  the order of integration and we have used the following equalities: $\int_{\Bbb{R}}e^{-ik(t-t_1)}\,dk=2\pi\delta(t-t_1)$ and  $\int_0^t g(t_1)\delta(t-t_1)\,dt_1=\frac{1}{2}g(t)$, for any test function $g(t)$.
Then, its solution can be written as
\be a(t)=\left(a-i\lambda \int_{\Bbb{R}}dk\eta(k,t)b(k)\right)e^{-(i\omega+\pi\lambda^2)t},
\label{sr6}\en
where $\eta(k,t)=\frac{1}{\rho(k)}\left(e^{\rho(k)t}-1\right)$ and $\rho(k)=i(\omega-k)+\pi\lambda^2$. Using complex contour integration it is possible to check that $[a(t),a^\dagger(t)]=\1$ for all $t$: this means that the natural decay of $a(t)$ described in (\ref{sr6}) is balanced by an opposite reservoir contribution. This is expected because of the existence of the integral of motion $\mathfrak N$, and it is crucial since it is a measure of the fact that the time evolution is unitarily implemented as it must be since $H$ is self-adjoint, even if $a(t)$ goes to zero with $t$. Let us now consider a state over $\ST\otimes\R$, $\left<\,X_\ST\otimes X_\R\right>=\left<\varphi_{n_a},X_\ST\varphi_{n_a}\right>\,\left<\, X_\R\right>_\R$, in which $X_\ST$ and $X_\R$ are, respectively, operators of the system and of the reservoir, $\varphi_{n_a}$ is the number eigenstate of $\hat n_a$, and $<\,>_\R$ is a state of the reservoir, which is assumed to satisfy $\left<\, b^\dagger(k)b(q)\right>_\R=n_b(k)\delta(k-q)$. This is a standard choice, see for instance \cite{barrad}, which naturally extends to $\R$ the choice we have made for $\ST$. Then, if we take $n_b(k)$ to be constant in $k$ we get, calling $n_a(t)=<\hat n_a(t)>=<a^\dagger(t)a(t)>$,
\be
n_a(t)=n_a\,e^{-2\lambda^2\pi t}+n_b\left(1-e^{-2\lambda^2\pi t}\right),
\label{sr7}\en
which goes to $n_b$ as $t\rightarrow\infty$. Hence, if $0\leq n_b<n_a$, the value of $n_a(t)$ decreases with time. If, on the other way, $n_b>n_a$, then the value of $n_a(t)$ increases for large $t$. This is the exponential rule which, as we have shown before, cannot be deduced if $\R$ has not an infinite number of degrees of freedom.

It might be interesting to remark that the continuous reservoir could be replaced by a discrete one, in which we have again an infinite number of actors, but they are labeled by a discrete index. In this case, to obtain a Dirac delta, rather than the integral $\int_{\Bbb{R}}e^{-ik(t-t_1)}\,dk=2\pi\delta(t-t_1)$, we should use the Poisson summation formula, which we  write here as $\sum_{n\in\Bbb{Z}}e^{inxc}=\frac{2\pi}{|c|}\sum_{n\in\Bbb{Z}}\delta\left(x-n\frac{2\pi}{c}\right)$. However, we will not consider this possibility here.

\end{document}